\documentclass{emulateapj}
\usepackage{natbib,graphicx,amsmath,amsthm}
\def\refnew#1{(\ref{#1})}

\def\erg{\, \rm erg}

\def\s{\, \rm s}
\def\km{\, \rm km}
\def\cm{\, \rm cm}
\def\m{\, \rm m}

\def\g{\rm g}

\usepackage{ulem}  
\usepackage{color}

\begin{document} 

\title{\mbox{ Planetesimals in Debris Disks of Sun-like Stars}}

\author{Andrew Shannon \& Yanqin Wu} 

\affiliation{Department of Astronomy \& Astrophysics, University of Toronto, Canada, M5S 3H4}



\begin{abstract}
  Observations of dusty debris disks can be used to test theories of
  planetesimal coagulation. Planetesimals of sizes up to a couple
  thousand kms are embedded in these disks and their mutual collisions
  generate the small dust grains that are observed.  The dust
  luminosities, when combined with information on the dust spatial
  extent and the system age, can be used to infer initial masses in
  the planetesimal belts. Carrying out such a procedure for a sample
  of debris disks around Sun-like stars, we
  reach the following two conclusions. First, if we assume that
  colliding planetesimals satisfy a primordial size spectrum of the
  form $dn/ds \propto s^{-q}$, observed disks strongly favor a value
  of $q$ between $3.5$ and $4$, while both current theoretical
    expectations and statistics of Kuiper belt objects favor a
    somewhat larger value.  Second, number densities of planetesimals
  are two to three orders of magnitude higher in detected disks than
  in the Kuiper belt, for comparably-sized objects. This is a surprise
  for the coagulation models. It would require a similar increase in
  the solid surface density of the primordial disk over that of the Minimum Mass Solar Nebula,
  which is unreasonable.  Both of our conclusions are driven by the need to explain the
  presence of bright debris disks at a few Gyrs of age.  

\end{abstract}


\section{Introduction}
\label{sec:introduction}

Dusty disks made up of rocky and icy debris have been observed around
other stars, both in reflected optical light \citep{Smith:1984} and in
long wavelength thermal radiation \citep{Aumann}. Multiple surveys
have reported that a significant fraction of main-sequence stars
harbor detectable infrared excesses: $\sim 15\%$ solar-type stars
\citep{Trillingetal:2008,Lawler}, and $\sim 30\%$ for A-stars
\citep{Suetal06}.  The infrared luminosity, when compared to the
luminosity of the central star, ranges from $\sim 10^{-5}$ to $\sim
10^{-3}$.  In contrast, the fractional dust luminosity from the Kuiper
belt is estimated to be $\sim 10^{-7}$ \citep{Teplitzetal:1999} and
remains undetected.

The observed excess luminosities arise primarily from small ($\sim \mu
m - mm$) dust grains. Due to their short survival time \citep{Pawel}, these grains are
believed to be continuously produced by collisions between large parent
bodies (`planetesimals'). These planetesimals, analogous to the Kuiper
belt objects in our own system, are in turn left-overs from the epoch
of planet formation.

In this article, we describe how we can use debris disks to test
theories of planetesimal formation. We first focus our attention on
the primordial size spectrum of planetesimals, often characterized by a
single power-law, $dn/ds \propto s^{-q}$, where $s$ is the size.  In
the following, we briefly summarize theoretical understandings and
observational evidences for the value of $q$.

The conventional picture of planetesimal formation is composed of a
number of steps.  The formation of the first generation planetesimals
is not yet well-understood and is an area of active research
\citep[see, e. g.][]{Youdin:2002,Dominik:2007,
  Johansen:2007,Garaud:2007}. If these are sufficiently massive,
gravity dominates their subsequent growth \citep{Weidenschilling}. At
first, objects grow in an orderly fashion, where collisions and
conglomerations occur at rates that are proportional to their
geometric cross sections.  But when these bodies become so massive
that the effect of gravitational focusing becomes significant,
run-away growth commences where the largest bodies accrete small
planetesimals at the highest rate and quickly distance themselves from
their former peers \citep{WetherillStewart,Kokubo96}. The run-away
phase is succeeded by the oligarchic phase where individual large
bodies are responsible for stirring the small bodies that they accrete
\citep{KokuboIda,Kokubo:1998}.  At the end of these steps, an entire
size spectrum of planetesimals are produced.  This is the `primordial
spectrum'.

During the run-away phase, N-body simulations have typically produced
a slope of $q \sim 6$ \citep{Kokubo96,Morishima:2008}.
This slope is naturally explained if there is energy equi-partition
among planetesimals of different sizes \citep{Makino:1998}. Moreover,
one expects that the distribution becomes shallower (smaller $q$) if
larger planetesimals have higher kinetic energies. This indeed occurs
during the oligarchic phase when all small and intermediate-sized
planetesimals are stirred to the same velocity dispersion. The value of
$q$ is then reduced to $\approx 4$ \citep{Morishima:2008}.

Using particles-in-a-box simulations and later hybrid simulations,
\citet{KenyonandLuu:1999, Kenyon:2004, 2008ApJS..179..451K} followed
the growth of planetesimals. They also found that $q$ decreases with
time after the run-away phase, finishing up with $3.75 \leq q \leq 4.5$
for planetesimals of sizes between $10$ and $1000$~kms. Recently,
\citet{2011ApJ...728...68S} argued analytically that a $q=4$
spectrum is the natural outcome of conglomeration.

Observational constraints on the value of $q$ currently come
exclusively from counting large Kuiper belt objects.  Kuiper belt
objects larger than about $30-50$~kms are commonly believed to be
primordial. Collision timescales for these bodies well exceed that of
the Solar system age
\citep{1997Icar..125...50D,2010AJ....139.1499B}. The size distribution
for these bodies can be probed by present-day surveys.  Published
values for $q$ are scattered: $q =4.0_{-0.6}^{+0.5}$
\citep{Trujillo:2001}, $q = 4.25 \pm 0.25$ \citep{Fraser:2008b}, $q =
4.5\pm 0.4$ \citep{2009AJ....137...72F} and $q = 4.5_{-0.5}^{+1.0}$
\citep{2008AJ....136...83F}. This scatter may be intrinsic and reflect
both the different size ranges and the different dynamical populations
emphasized by various surveys
\citep{Bernstein,2006P&SS...54..243D,FraserBrown}.  For bodies
smaller than $\sim 30$~kms, the size distribution adopts a shallower
power-law \citep{Bernstein,2008AJ....136...83F,2009Natur.462..895S}.
This break in the power-law index has been argued to be due to
collisional erosion \citep{PanSari}, but a different opinion has
surfaced \citep{morbidelli}.

So at least for the value of $q$, current coagulation models appear to
be vindicated by the observations. These models enjoy a further
success.  In the Kuiper belt region, the solid mass of the so-called
Minimum Mass Solar Nebula is $\sim 10~M_\oplus$
\citep{Hayashi,Weidenschilling:1977}, while the mass in large Kuiper
belt objects is estimated to be $\lesssim 0.1 M_\oplus$ \citep[see,
e.g.][]{Gladman,Bernstein}. This large difference, however, is
explained by current models where the formation of large planetesimals
has a very low efficiency
\citep{2006AJ....131.2737B,2011ApJ...728...68S}.

With these two remarkable concordances, one wonders if debris disks
will ever tell us anything new and unexpected.  Furthermore, every
debris disk likely has a different initial condition and evolves in a
different dynamical environment. For instance, dynamical interactions
with Neptune or other planets may have qualitatively affected the
evolution of the Kuiper belt \citep{Levison:2008}.  It seems
difficult, therefore, to extract any universal truth about the
formation process from these disparate objects.

However, based only on a modest sample of debris disks, we argue in
this paper that there is already a serious issue in current
coagulation models.

To achieve this, we first construct a simple collisional model (\S
\ref{sec:luminosityevolution}) to compare against the set of debris
disks reported in \citet{Hillenbrand:2008}.  Our collisional model
does not differ in essence from previous works
\citep{Krivov05,Wyattetal:2007,Lohne:2008}, but we interpret the
observations in a new way. This allows us to measure the value of $q$
as well as the initial masses of planetesimal belts (\S
\ref{sec:results}). The latter result challenges the current models of
planetesimal formation (\ref{sec:discussions}). We summarize in \S
\ref{sec:summary}.

\section{Model: Luminosity Evolution of a Debris Disk}
\label{sec:2}
\label{sec:luminosityevolution}

The debris phase commences when eccentricities of the primordial
planetesimals are further increased so that they no longer coalesce at
encounter, but are instead broken into
fragments.\footnote{\citet{2008ApJS..179..451K} find that
  fragmentation begins once Pluto-sized bodies form.}  In this phase,
the smallest primordial planetesimals enter into a collisional cascade
first, followed by progressively larger bodies. During the collisional
cascade, a primordial body is broken down into smaller and smaller
fragments until all its mass ends up in small grains. The small grains
may spiral in towards the star due to Poynting-Robertson drag, as
happens in the Solar system, or, be ground down by frequent collisions
to sizes so small that they are promptly removed by radiation
pressure, as happens in bright debris disks \citep{Wyatt:2005}.

\subsection{Debris Rings}

We model the debris disk as a single, azimuthally smooth ring composed
of planetesimals of different sizes. The ring is centered at a
semi-major axis $a$ with a full radial width of $\Delta a$ and a
constant surface density.  We take $\Delta a/a = 0.1 \ll 1$ as our
standard input.  
This is motivated by the following observations.  Spatially resolved
debris disks often appear as narrow rings. Examples are, $\Delta a/a$
$\sim 0.1$ for AU Microscopii \citep{2007ApJ...670..536F}, $\sim 0.5$
for HD 10647 \citep{2010A&A...518L.132L}, $\sim 0.3$ for HD 92945
\citep{2007lyot.confE..46G}, $\sim 0.3$ for HD 139664
\citep{2006ApJ...637L..57K}, $\sim 0.2$ for HD 207129,
\citep{2010AJ....140.1051K}, $\sim 0.5$ for $\epsilon$ Eridani
\citep{2000MNRAS.314..702D}, $\sim 0.1$ for Fomalhaut
\citep{Kalas:2005}, $\sim 0.2$ for Vega \citep{2005ApJ...628..487S}.
Similarly, unresolved disks often exhibit spectral energy distribution
that is well fit by a single temperature blackbody
\citep{Hillenbrand:2008, 2010A&A...518A..40N,Moor}. This ring-like
topology also show up in our own Solar system, hence the name the
asteroid ``belt'' and the Kuiper  ``belt''.

\subsection{Initial Size Distribution of the Planetesimals}
\label{subsubsec:dnds}

We adopt the following power-law forms for the initial size
distributions,
\begin{equation}
\left. \frac{dn}{ds}\right|_{t=0} \propto 
\begin{cases}
s^{-q_3} \quad \quad s_{\rm small} < s < s_{\rm big} , \\
s^{-q_1} \quad \quad s_{\rm min} < s < s_{\rm small} .\\
\end{cases}
\label{eq:time0}
\end{equation} 
The index $q_3$ is the primordial size index for large bodies, like
one that arises out of conglomeration models.  Previous studies of
collisional debris disks have taken this value to be a given, in fact
it is commonly set to be the power law one expects from collisional
equilibrium \citep{Krivov05,Krivov06,Wyattetal:2007,Lohne:2008}.
In contrast, in this contribution we use the observed sample to
measure this value.

In equation \refnew{eq:time0}, $s_{\rm big}$ is the size of the biggest
planetesimals, $s_{\rm min}$ the smallest. The intermediate size
$s_{\rm small}$ is introduced for the purpose of mass accounting: the
original mass counts only those between $s_{\rm big}$ and $s_{\rm
  small}$,
\begin{equation}
M_0 = \int_{s_{\rm small}}^{s_{\rm{big}}} \frac {4\pi}{3}\rho s^3\, n_3s^{-q_3}\,ds.
\label{eq:m0defined}
\end{equation}
While $s_{\rm min}$ is naturally taken to be the size at which
radiation pressure unbinds dust grains from the star ( $\sim \mu m$
for a Sun-like star), we discuss our choice for $s_{\rm big}$ and
$s_{\rm small}$ below.

Motivated by the observational and numerical results discussed in \S
\ref{sec:introduction}, we investigate values of $q_3$ between $3.5$
and $5$.  The value $q_3 = 4$ has the special property that mass is
distributed equally among all logarithmic size ranges, while masses in
systems with $q_3 > 4$ diverge toward the small end. The intermediate
size $s_{\rm small}$ is introduced, partly to avoid dealing with this
divergence.  For sizes below $s_{\rm small}$, we assume that
collisions have set up an equilibrium power law with index $q_1$ (see
Appendix).  So, the intermediate size $s_{\rm small}$ can also be
interpreted as the collisional break size at time zero. For our study,
we set $s_{\rm small} = 100$ m.
For our typical disks, we find that, within a few million years,
collisional equilibrium is established for bodies up to sizes $\sim 1$
km. So the choice of $s_{\rm small}$ is not important for late time
evolution.


The choice of size for the largest bodies, $s_{\rm big}$, deserves
some discussion, as it affects the qualitative character of the
evolution.  As a collisional cascade progresses, bodies of larger and
larger sizes come into collisional equilibrium, opening up fresh mass
reserve to produce the small particles. Once the largest bodies enter
into collisional equilibrium, the dust production rate decays with
time as $L_{\rm IR} \propto t^{-1}$ \citep{2007ApJ...658..569W}.
Two previous studies \citep{Wyattetal:2007,Lohne:2008} have adopted
sizes for the largest bodies of $s_{\rm big} = 30$ and $74$ km,
respectively.  For some of their disks, the largest bodies can enter
collision equilibrium during the lifetime of the system.

Both Kuiper belt observations and numerical studies of coagulation
favor a largest size of $\sim 1000$~km.  The largest object yet found
in the Kuiper Belt, (136199) Eris, has a radius of $1200 \pm 50$~km
\citep{Brown:2006}. In the simulations of \citet{Kenyon:2004b},
coagulation of planetesimals at 30 - 150 AU produces bodies as large
as $1000$ - $3000$ km.  When the largest bodies reach this size,
self-stirring increases the velocity dispersion and collisions become
destructive rather than conglomerating.

Therefore, we adopt a maximum body size of $1000$ km in our study. Our
quoted masses reflect this choice of $s_{\rm big}$.  Our largest
bodies never enter into collisional equilibrium. If this assumption
turns out to be erroneous, namely, $s_{\rm big}$ is much smaller and
enters into collisional cascade within system lifetime, our model
would underestimate the initial masses for old disks.  As a result, we
would overestimate the value for $q_3$.

\subsection{Collisions}
\label{subsubsec:collision}

We only consider collisions that are catastrophically destructive.  A
catastrophic collision is defined as one that removes at least $50\%$
of the mass of the primary body.  In so doing, we have implicitly
assumed that both cratering collisions and conglomerating collisions
are unimportant.  When a destructive collision occurs, the total mass
(bullet plus target) is redistributed to all smaller sizes according
to $dn/ds \propto s^{-4}$. This choice is somewhat arbitrary and we
have confirmed that modifying it (within reasonable bounds) does not
change our results.

We do not model evolution of the orbital dynamics as bodies
collide. This is justified by the discussions in \S
\ref{subsec:eccentricity}.

Let the chance of collisions between two bodies of sizes $s$ and
$s^{\prime}$ be,
\begin{equation}
f_{\rm{collision}} = \frac{\pi \left(s + s^{\prime}\right)^2}
{2 \pi a \Delta a \, t_{\rm{orb}}},
\label{eq:fcol}
\end{equation}
Here, $2\pi a \Delta a$ is the surface area spanned by the debris ring
in the orbital plane, and $t_{\rm orb}$ is the orbital period.
Gravitational focusing is negligible for the high random velocities
we consider here.  The typical encounter velocity, for particles with
eccentricity $e$ and inclination $i$, is \citep{Wetherill:1993}
\begin{equation}
v_{\rm{col}} = \sqrt{1.25e^2+i^2}\, v_{\rm{kep}},
\label{eq:vimpact}
\end{equation}
where $v_{\rm kep}$ is the local Keplerian velocity.  We adopt $i
\approx e/2$ so $v_{\rm col} \approx 1.32\, e \, v_{\rm kep}$. As
argued in \S \ref{subsec:eccentricity}, it is reasonable to assume a
constant eccentricity (and inclination) for all bodies. We take a
value of $e=0.1$ as the standard input, and discuss this
  assumption in \S \ref{sec:discussions}.

We denote the specific impact energy required to catastrophically
disrupt a body (target) as ${Q}^*$. The scaling of $Q^*$ with the size
of the target depends on whether its strength is dominated by material
cohesion or self-gravity. We adopt the following form
\citep{BenzAndAsphaug},
\begin{equation}
{Q}^*~=~A\left(\frac{s}{1~\mbox{cm}}\right)^\alpha+B\rho\left(\frac{s}{1~\mbox{cm}}\right)^\beta
\label{BenzAndAsphaug}
\end{equation}
where $\rho$ is the bulk density which we take to be $2.5
\rm{g}/\cm^3$. The first term on the right-hand-side describes the
internal strength limit, important for small bodies, while the second
term the self-gravity limit, important for larger bodies.

The strength law sets the size of the smallest bullets required to
destroy a target.  Since these are also the most numerous, they
determine the downward conversion rate of mass during a collisional
cascade. As such, the power indexes in the strength law directly
determine the size spectrum at collisional equilibrium. For a strength
law of the form ${Q}^* \propto s^c$, the equilibrium size spectrum is
$dn/ds \propto s^{-q}$, with \citep{1997Icar..130..140D}:
\begin{equation}
q=(21+c)/(6+c).
\label{eq:whatisq}
\end{equation}
The famous Dohnanyi-law \citep{Dohnanyi:1969}, $dn/ds \propto
s^{-3.5}$, obtains from $c = 0$.

\begin{figure}[t]
\begin{center}
  \includegraphics[scale=.65, trim = 0 0 0 0, clip]{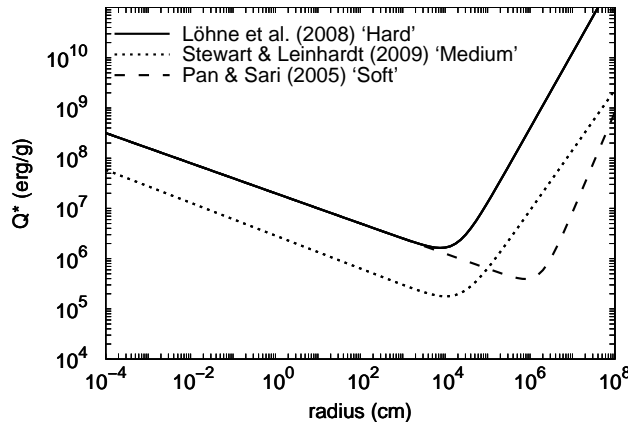}
  \caption{ Prescriptions for specific strength from \citet{Lohne:2008},
    \citet{PanSari} and \citet{2009ApJ...691L.133S}, plotted here as
    functions of target sizes. We insert an impact velocity of $500
    \m/\s$ to evaluate the last prescription. Strength of small bodies
    are dominated by material cohesion, while that of larger bodies by
    self-gravity.  Transitions between the two limits occur around
    $100$ m (the hard and the medium laws) or around $10$ km (the soft
    law).  Strength for bodies smaller than $1 \cm$ are extrapolations
as both laboratory and numerical experiments only concern bodies of
larger sizes.}
\label{fig:Qstarcomp}
\end{center}
\end{figure}

The value and form for $Q^*$ are notoriously difficult to assess. It
depends on, among other factors, material composition, porosity and
impact velocity. A number of computations and compilations have
appeared in the literature. We select three representative
formulations for our study (Fig. \ref{fig:Qstarcomp}).

Based on a variety of experimental data and SPH simulations,
\citet{Krivov05,Lohne:2008} advocated the following choices, $A = 2
\times 10^7 \erg/\g$, $\alpha = -0.3$, $B = 0.158$, $\beta = 1.5$. We
call this the 'hard' strength law.  In this case, the collision
spectrum satisfies $q \approx 3.6$ and $3.0$, in the strength and
gravity regimes respectively.

Based on energy conservation, \citet{PanSari} calculated a destruction
threshold for bodies that have zero internal strength and obtained $B
= 3.3 \times 10^{-8}$, $\beta = 2$. So bodies at $100 \km$ is weaker
by a factor $\sim 1000$ than their counterparts in the
\citet{Krivov05} formulation. We refer to this as the 'soft' strength
law. A softer strength implies smaller bullets and therefore more
frequent destruction of the targets. \citet{PanSari} did not consider
smaller bodies that are strength bound. We adopt $A = 2 \times 10^{7}
\erg/g$ and $\alpha = -0.3$ in this range to complete the soft
prescription.

\cite{2009ApJ...691L.133S} proposed a strength law that depends on
impact velocity,
\begin{equation}
Q^* = \left(500\, s^{-0.33}+10^{-4}\,  s^{1.2}\right)v_{\rm col}^{0.8}, 
\label{eq:sllaw}
\end{equation}
For a typical velocity $v_{\rm col}= 500$m/s and for bodies
  greater than 1km, this gives rise to a strength law that falls
in-between that of the hard and the soft case. We call this the medium
strength law. Note that this strength law is much weaker than
  the other two for small bodies.

For the strength laws we consider, transitions from material strength
domination to self-gravity domination occur at size $s \approx s_1$,
with $s_1$ ranging between $100$ m (the hard and the medium laws) and
$10$ km (the soft law).

\subsection{Luminosity Evolution}

\begin{figure}[]
\begin{center}
  \includegraphics[scale=.65]{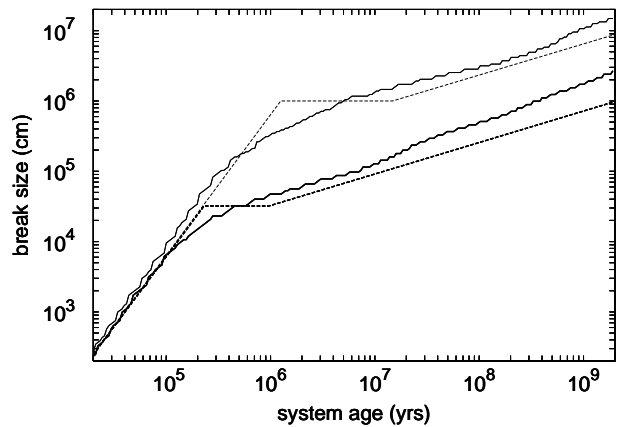}
  \caption{Time evolution of the break-size in a model system, with
    $M_0 = 12.4 M_{\oplus}$, $q_3 = 4.0$, $e = 0.1$, $a = 31\rm{AU}$,
    and $\Delta a/a = 0.1$.  Here, break-size is defined as the size
    at which all bodies initially at that size have encountered of
    order one destructive collision. Break size increases with time
    monotonically as larger bodies enter into collisional cascade.
    The numerical results are shown as solid curves, while the
    analytical scaling relations (see Appendix) are plotted as dashed
    lines.  The bends in the curves occur at $s \approx s_1$, i.e.,
    sizes for which material cohesion and gravity binding are
    comparable. The set of thick curves are for the case of hard
    material strength, while the thin lines for soft strength. }
\label{fig:s2}
\end{center}
\end{figure}

\begin{figure}[t]
\begin{center}
\includegraphics[scale=.65, trim = 0 0 0 0, clip]{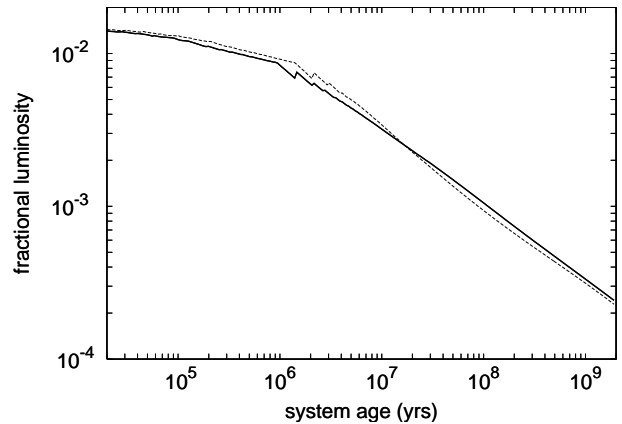} 
\caption{Evolution of fractional luminosity, $L_{\rm IR}/L_*$, for the
  system in Fig. \ref{fig:s2}. The thick line is obtained using the
  hard strength law, and the thin line the soft one. The evolution
  proceeds in two stages: the flatter early stage when collisional
  cascade only involves small bodies that are bound by material
  cohesion; and a steeper later stage where bodies bound by
  self-gravity enter the cascade.  At late times, fractional
  luminosity decays as $t^{-0.5}$ (eq. \refnew{eq:ftime}). While
  break-sizes differ for the two adopted strength laws
  (Fig. \ref{fig:s2}), this appears to have little influence on the
  overall luminosity.  }
\label{fig:flHD1}
\end{center}
\end{figure}

The planetesimal disk, starting from an initial disk mass of $M_0$,
and an initial size spectrum (eq. \ref{eq:time0}), is numerically
collided and ground down. We divide the particles between $s_{\rm
  small}$ and $s_{\rm big}$ into $500$ equal logarithmic size bins.
The time-step for the simulations is adaptively set so that over one
time-step, the maximum mass gain (from larger bodies) or loss (to
smaller bodies) per bin falls below $5\%$.  The net mass change is
substantially smaller than this due to the cancellation between gain
and loss.

We calculate the fractional brightness of the dust disk, $L_{IR}/L_*$,
by integrating the geometrical cross section over all grains. This
assumes that grains are perfect absorbers at the optical and can emit
efficiently in the infrared.

An example of such a calculation is reported in Figs. \ref{fig:s2} \&
\ref{fig:flHD1}. To understand these results, a simple analytical
model (see Appendix) is introduced. Scaling relations obtained using
this analytical model compares well with our numerical results.

Fig. \ref{fig:s2} shows that, with time, larger and larger
planetesimals enter into collisional cascade. Within a million years or
so, the cascade has advanced to size of order one kilometer. Beyond
this time, bodies bound by self-gravity can be gradually eroded. By 1
Gyrs, bodies with sizes $10-100$ kms may be affected. The exact value
depends on the strength law. The dust luminosity is related to the
dust mass, which is in turn related to the dust production rate.  The
dust production rate, on the other hand, is simply the primordial mass
stored at the break-size divided by the system age.  If the primordial
spectrum is such that a large amount of mass is piled at the large
end, debris disks would not exhibit significant fading even up to a
few billion years.

Fig. \ref{fig:flHD1} shows that dust luminosity $L_{\rm IR}/L_*
\propto t^{-0.5}$ for $q_3 = 4$, consistent with equation
\refnew{eq:frac2}. That same equation also demonstrates that the value
of $B$, strength constant for bodies bound by self-gravity, affects
the luminosity only minorly. This is born out by
results shown in Fig. \ref{fig:flHD1}.

\begin{figure}[t]
\begin{center}
\includegraphics[scale=.65,trim=0 0 0 0,clip]{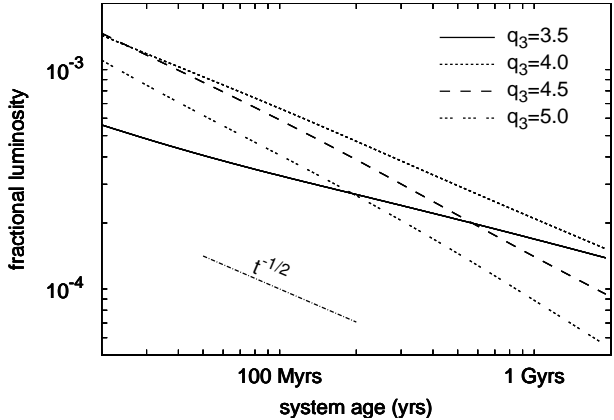}
\caption{Luminosity evolution for disks with different $q_3$ but the
  same initial mass ($8 M_\oplus$). Systems with a steeper primordial
  size spectrum (larger $q_3$) exhibit a more pronounced decline of
  luminosity with time, since at a given time, a bigger fraction of
  their mass reservoir has been depleted.  Systems with shallower
  $q_3$ (e.g., $q_3 = 3.5$), on the other hand, are initially dimmer
  due to the relative shortage of smaller rocks, but eventually
  outshine the higher $q_3$ disks as they can hold on to their mass
  reservoir for longer.  The luminosity decay of observed disks that
  span a large range of ages can thus be used to infer the value of
  $q_3$. All other parameters here are similar to those used in
  Fig. \ref{fig:flHD1} and we adopt the hard strength law.}
\label{fig:varyq}
\end{center}
\end{figure}

An important result on which we base our later analysis is shown in
Fig. \ref{fig:varyq}. Luminosity evolution for disks with the same
initial mass but different $q_3$ are depicted. As
eq. \refnew{eq:ftime} predicts, $L \propto t^{(q_3-3)/(2-q_3)}$.  If
$q_3$ is shallow (e.g. $q_3 \leq 4$), most of the initial mass is
deposited at the largest planetesimals. This mass reservoir is harder
to reach by collision and allows the disk to remain brighter at later
times. In comparison, disks with a steeper $q_3$ decay faster.

If one observes a collection of debris disk all at the same age,
intrinsic scatter in, e.g., initial masses, makes it impossible to
differentiate between models of different $q_3$. However, a collection
of disks with a large age spread can be used to constrain $q_3$. This
we proceed to demonstrate.

\section{Observed Ensemble}
\label{sec:ensemble}

Several debris disks surveys have been carried out \citep[see,
e.g.][]{Suetal06,Trillingetal:2008,Lawler, Moor}.  The sample of most
interest to us is that reported in \citet{Hillenbrand:2008}.  Together
with updates in \citet{2009ApJS..181..197C}, \citet{Hillenbrand:2008}
presented a collection of debris disks around F/G/K type stars,
obtained as part of the Spitzer program on Formation and Evolution of
Planetary Systems (FEPS). This sample is unique in that both the
stellar age and the radial distance of the dust ring are determined:
isochrone fitting provides the age for the host stars (spanning from
$\sim 10^{7}$~years to a few $10^{9}$~years), while multi-band
photometry and spectral energy fitting yield the semi-major axis of
the dust ring.  Together with fractional luminosity of the dust belt,
these provide the most important constraints to infer the primordial
properties of parent planetesimals.

To obtain the blow-out size ($s_{\rm min}$) for each system, we take
luminosity values for the central stars as given in
\citep{Hillenbrand:2008}, and we assign stellar masses by assuming
that $M_* \propto L_*^{1/3}$, as appropriate for solar type main-sequence
stars.

Out of the $31$ disks listed in \citet{Hillenbrand:2008}, we focus
only on a sub-sample of 13 disks that appear radially unextended and
are around main-sequence stars. In \citet{Hillenbrand:2008}, emission
from each disk is initially fitted with a single temperature blackbody
(a ring).  If agreement between the ${24 \mu \rm{m}}/{33\mu \rm{m}}$
fit and the ${33 \mu \rm{m}}/{70 \mu \rm{m}}$ fit is poor,
they argue that the disk is likely radially extended and fit the data
instead with two radial components.
Since our numerical model is a one-zone model, we find that including
the extended sources into our analysis causes significant scatter in
the results. This leads us to discard them for the current
analysis. We have excluded HD 191089 from our sample.  Its fluxes in
13 $\mu$m and 33 $\mu$m are not measured, and cannot be reliably
identified as an unextended source. In all, we are left with 13
sources.

\begin{figure}[]
\centering
  \includegraphics[scale=.65]{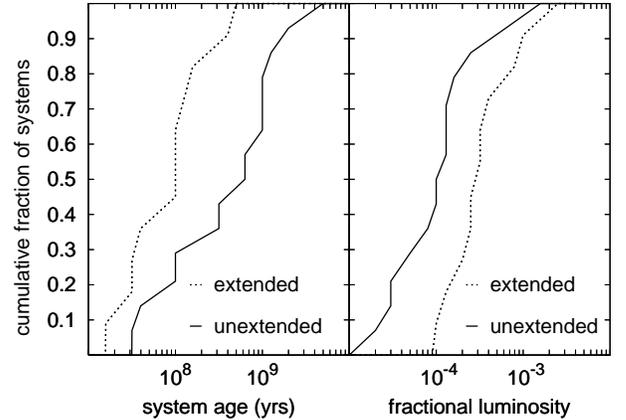}
  \caption{Cumulative distribution of stellar age (left panel) and
    dust luminosity (right panel) for the \citet{Hillenbrand:2008}
    sample. The extended systems (solid curve) tend to be younger and
    brighter than the unextended systems (dashed curves). }
\label{fig:compage}
\end{figure}

It is interesting to note that most of the extended sources are
relatively young, all younger than a few hundred million years. In
contrast, the unextended sources have a larger age spread, lasting
till a few billion years (Fig. \ref{fig:compage}).  All systems may be
born with more than one debris rings, but after a sufficiently long
time, only the outermost ring, which has the longest erosion
timescale, remains shining. The extended system are also brighter than
the average, likely related to their relative youth.

\begin{figure*}
\centering
\includegraphics[scale=1.1]{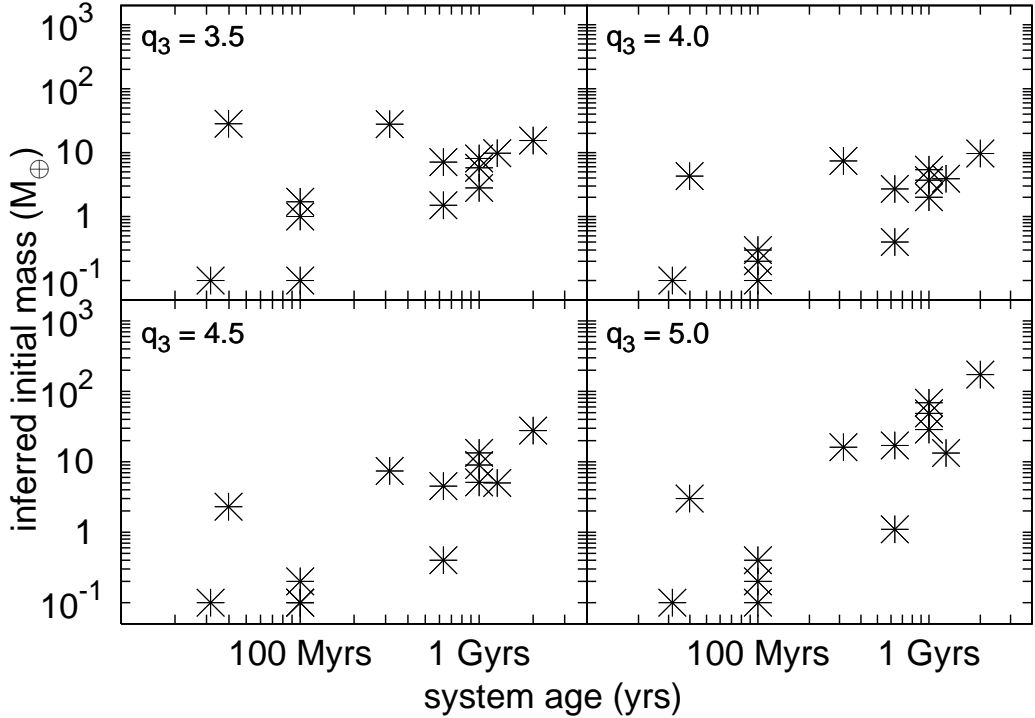}
\caption{Inferred disk initial masses, plotted against system ages,
  for the unextended systems in \citet{Hillenbrand:2008}. The four
  panels present four different choices of $q_3$. The other parameters
  chosen are $e = 0.1$, $s_{\rm small} = 10^{4} \rm{cm}$,~and $\Delta
  a/a = 0.1$.  A value of $q_3 \in [3.5 ,4.0]$ is preferred: the upper
  envelopes for the disk mass remain constant at all ages in the two
  top plots.  Models with higher $q_3$ are excluded as they require a
  rising upper envelope. In addition, the $q_3 = 5$ model requires
  unphysically large disk masses for very old disks.  }
\label{fig:main}
\end{figure*}

\begin{figure*}
\centering
\includegraphics[scale=1.1]{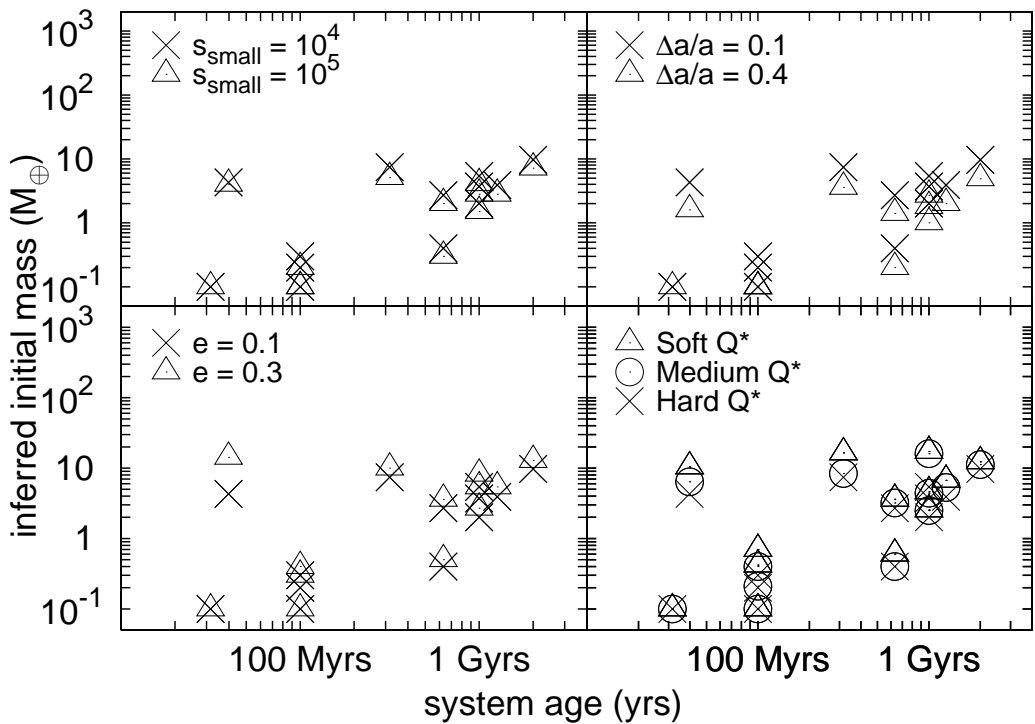}
\caption{ Similar to Fig. \ref{fig:main} except $q_3$ is fixed at $4$
  and a number of parameters are varied to test how the inferred
  initial masses depend on them.  Specific values for the inferred
  mass may change, when the small end of the primordial size spectrum
  ($s_{\rm small}$, top-left), the eccentricity of particles
  (bottom-left), the fractional width of the debris ring (top-right),
  and the adopted strength law (bottom-right) are varied for all
  systems. However, the important indicator for our study, the upper
  envelope of the masses as a function of system age, remains flat.
  So the conclusion that $q_3 \sim 4$ remains valid. }
\label{fig:9to13}
\end{figure*}

\section{The Primordial Size Spectrum Revealed}
\label{sec:results}

We have a simple strategy. Knowing the luminosity, the age and the
semi-major axis of each debris ring, we use our collisional model to
backtrack the evolution to infer its initial mass in the planetesimal
belt. These initial masses, when plotted against system ages, should
show a spread. One expects this spread to be constant across all ages,
as disks formed at different cosmic times likely have the same
distribution of disk properties. This property could be used to test
model assumptions. However, using the spread is difficult due to
selection effects. For instance, low mass disks may become too dim at
late times to be observable. So we propose instead to study the upper
envelope of this spread. The upper envelope should be flat with age
for the correct model.  From our analytical scaling relations (see
Appendix), we find that the most important parameter in our model that
affects this mass slope is $q_3$, the power-law index in the
primordial size spectrum.

The results of such a procedure are shown in
Fig. \ref{fig:main}. Models with $q_3 = 4.5$ or greater appear to be
excluded by data, as they would require a rise of initial disk mass
with stellar ages. The reason behind this is transparent by studying
Fig.  \ref{fig:varyq}.  Models with $q_3 = 3.5$ and $4$ are compatible
with observations. Models with smaller $q_3$ lead to a
decreasing initial mass with system age and are excluded as
well.

Our model employs a number of other parameters, such as the
  radial position and extent of the debris ring, the dynamical
  excitation and break-up strength of the particles.  We have studied
the robustness of our results when these parameters are varied
(Fig. \ref{fig:9to13}).  As long as the values for these
  parameters remain constant over age, varying them do not affect our
  conclusion on $q_3$. The assumption that the dynamical excitation is
  constant over age is suspicious, in light of results from
  coagulation models showing that stirring by large plantesemals
  increases gradually eccentricities of the disk particles. This is
  discussed in \S \ref{sec:discussions}.

There is significant uncertainty in our conclusion due to the small
sample size. However, we argue that a larger sample may still not
favor models with, e.g., $q_3 = 5$. If $q_3 = 5$ (lower-right panel in
Fig. \ref{fig:varyq}), the system that remains easily detectable at 2
Gyrs of age requires an initial solid mass of $\sim 200 M_\oplus \sim
1 M_J $ in the planetesimal belt. The initial gas mass in such a belt
will be higher than the total disk mass of a typical T-Tauri star
($0.01 M_\odot$).

By focusing on dust luminosities, we are sensitive only to bodies that
lie below the break-size.  As seen in Fig. \ref{fig:s2}, break-size
marches up to few tens to a hundred kilometers by the end of a few
billion years, if the disk has a mass of $M_0 = 12.4 M_\oplus$.

\section{Discussions}
\label{sec:discussions}

\subsection{Coagulation Models vs. Debris Disks}
\label{subsec:compare}

\begin{figure}[t]
\centering
\includegraphics[scale=0.65]{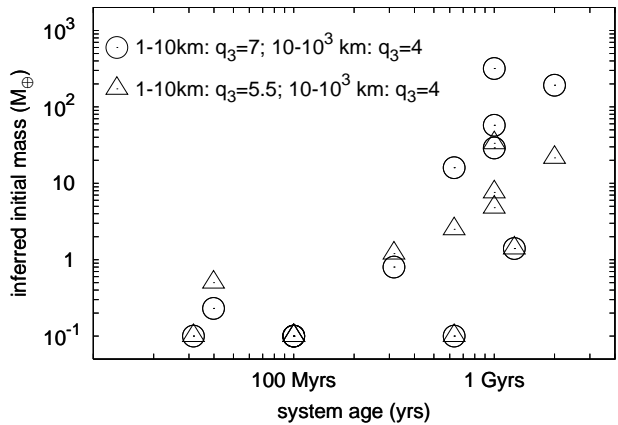}
\caption{Inferred initial masses for a broken power-law
  size-distribution. We investigate two particular forms, motivated by
  coagulation simulations by \citet{2008ApJS..179..451K} and \citet{2011ApJ...728...68S}, respectively. 
  Other parameters adopted are $e = 0.1$, $\Delta a/a = 0.1$, and the
  hard strength law. The inferred disk mass rises sharply with system
  age.  Moreover, to make old and bright systems, we require disk
  masses that approach the mass of Jupiter.}
\label{fig:kenyon}
\end{figure}

In our exercise, we have assumed a simple initial size distribution
(eq. \ref{eq:time0}), with all bodies larger than a few hundred meters
described by a single power law index $q_3$. We relax this assumption
here.

Simulations of planetesimal coagulation produce typically more
complicated size distributions. For example, \citet{2008ApJS..179..451K}
started their simulations with all bodies at $\leq 1$ km. After tens
of millions of years of growth, most of the mass still remains at or
below $1$ km, with only $\sim 8\%$ of the mass being accreted into
bodies $10$ km or larger, $\sim 6\%$ into bodies $100$~kms or larger,
and $\sim 3\%$ into bodies of order $1000$~kms.  We use a broken
power-law to replicate this kind of primordial spectrum. We set $q_3 =
5.5$~from $1$~km~to $10$~km, and $q_3 = 4$~from $10$~km~to $1000$~km.  Motivated by \citet{2011ApJ...728...68S},
we also consider a slightly different initial distribution with $q_3 =
7$~from $1$~km~to $10$~km, and $q_3 = 4$~from $10$~km~to $1000$~km.  Both sets of size spectrum deposit mass
mostly at the low end ($\leq 1$ km) and little at the large sizes.  As
expected, when initial masses are determined for different systems
(Fig. \ref{fig:kenyon}), we find that young systems require
exceedingly low initial masses, while old systems require unphysically
large initial masses.

If we follow the luminosity evolution of such a disk, we will see that
the disk flares brightly in the first tens of millions of years, due
to the large mass reservoir at the $1$ km-range. Then the luminosity
decays as $t^{-1/2}$ (as expected of a $q_3 = 4$ spectrum) but with a
low normalization -- most of the disk mass has been ground down in the
early stage and we are now left with but a scrap remnant of the
original. Conglomeration simulations typically find that only a small
fraction of the mass can be accreted to make large bodies, before viscous stirring effectively stalls the growth.
\citet{2011ApJ...728...68S} showed that the fraction in large
  bodies can only be of order $10^{-3}$ in Kuiper-belt-like
environments.

Does results in Fig. \ref{fig:kenyon} allow us to exclude current
conglomeration models? One possible caveat in our analysis is the
eccentricity.  We discuss this below.

\subsection{Eccentricity}
\label{subsec:eccentricity}

We assume a static, high eccentricity ($e=0.1$) for all systems at all
times. In realistic systems, eccentricities can be a function of time.

One possible cause of eccentricity evolution is collisional cooling.

Collisions dissipate energy, so collisional products have in average
lower velocity dispersion than their parent bodies. In a single
collision, two bodies with masses $m_1$ and $m_1^\prime$ (assume $m_1
\gg m_1^\prime$) impact with typical velocities\footnote{ Velocities
  here refer to the random component.}
\begin{equation}
v_1 \sim v_1^\prime \sim e_1 \, v_{\rm{kep}},
\end{equation}
where the subscript $1$ indicates that this is a first generation
collision in our counting.  Assuming that all collision debris fly
away from the collision site with the velocity of the center-of-mass,
i.e., all relative velocities in the center-of-mass frame is
dissipated during the collision, collisional cooling can be expressed
as
\begin{equation}
  v_2 \sim \frac{\sqrt{m_1^2v_1^2+m_1^{\prime 2}v_1^{\prime 2}}}{m_1+m_1^\prime} 
\sim v_1\left(1 - \frac{m_1^\prime}{2m_1}\right).
\end{equation}
So the closer in mass the two colliding bodies are, the more cooling
their debris experiences. If cooling dominates the eccentricity
evolution, we find that a minimum eccentricity of $e_1 = 0.13$ is
required (for the hard strength law, and $e_1 = 0.02$ for the medium
law) to allow collisional cascade to proceed all the way to micron
range. 

However, even if collisional cooling is severe, we argue that viscous
stirring by large planetesimals dominates the eccentricity evolution.
This is able to raise the eccentricity of collisional debris to values
comparable to that of their parents in a time shorter than a
collisional time. So the condition for a successful collisional
cascade is reduced to $e \geq 0.05$ for the hard strength law and $e
\geq 0.01$ for the medium strength law, i.e., the minimum random
motion necessary to break up the hardest grains (the smallest ones).
\footnote{This constraint can be reduced by a factor of unity when
  radiation pressure on small grains are considered
  \citep{Thebault09}.}

\begin{figure}[t]
\centering
  \includegraphics[scale=0.65]{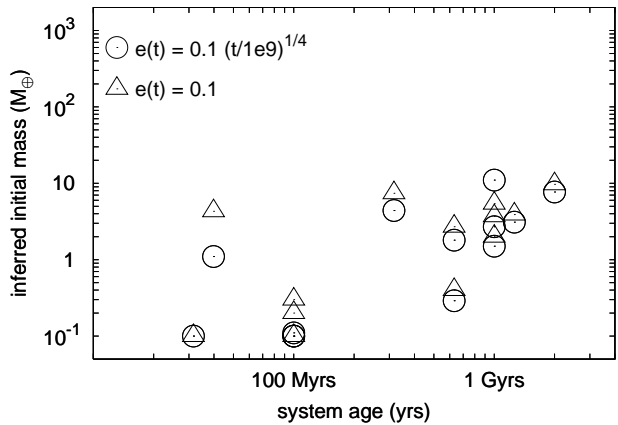}
\caption{Same as the top-right panel in Fig. \ref{fig:main} but
  instead of a constant eccentricity ($e=0.1$), here we assume that
  the eccentricity rises as $e(t) \approx 0.1 (t/10^9
  \rm{yrs})^{1/4}$. There is little difference to the inferred mass,
  and if anything, the data seems to argue that a $q_3= 4$ model
  slightly overestimate the value of $q_3$. So our conclusion that
  $q_3 \in [3.5,4]$ remains unchanged even considering eccentricity
  growth.  }
\label{fig:newfig}
\end{figure}

In fact, stirring is likely to gradually raise the eccentricity of all
bodies. Stirring by large bodies in the disk goes as
$e \propto t^{1/4}$
\citep[c.f.][]{GLS}.
In the simulations of \citet{2008ApJS..179..451K}, planetesimals are
continuously stirred by Pluto-like bodies, but they only reach $e\sim
0.1$ at about a Gyrs.\footnote{ An eccentricity of $e\sim 0.3$ at $40$
  AU corresponds to the surface escape velocity of
  Pluto. Planetesimals have to have a near-surface encounter before
  they can reach such a high eccentricity. This takes time.}
 Under such a scenario, the inferred initial disk mass is similar
 to the original result (Fig. \ref{fig:newfig}),
  but it is clear that we prefer the same range of values for $q_3$.

\section{Summary}
\label{sec:summary}

Using an ensemble of bright debris disks around Sun-like stars, we
have measured the size spectrum of their embedded planetesimals. We
parametrize the size spectrum as $dn/ds \propto s^{-q_3}$ and find
$q_3 \approx 3.5-4$, where $q_3 = 4$ corresponds to equal mass per
logarithmic decade.  The planetesimal sizes our technique probes lie
between a couple kms to $\sim 100 \km$.

While this size spectrum appears consistent with results of
coagulation simulations ($q_3 \sim 4$), there are two lines of
evidences that suggest problems in current coagulation models.

The first line of evidence is related to the inferre disk mass.  The
inferred initial masses for these bright disks are surprisingly
high. We find total masses reaching as high as $10
M_\oplus$.\footnote{This is for $q_3 = 4$, and even higher values are
  required if $q_3 = 3.5$.} This is comparable to the total solid mass
in the Kuiper belt region of Minimum Mass Solar Nebula model, and
about a factor of $100$ higher than the mass in large Kuiper belt
objects.  Current coagulation models require an MMSN-like total mass
to produce the observed density of large Kuiper belt objects. If the
same inefficiency persists for our disks, one would require a total
disk mass of $\sim 100$ MMSN to produce those embedded planetesimals.
This is difficult to imagine.

The second line of evidence regards the size spectrum.  We experiment
with size distributions that arise from coagulation simulations. We
find that these distributions could not reproduce the luminosity
distribution of the observed disks.  Current coagulation models are
highly inefficient in making large planetesimals.  So most of the mass
remains at where they started, presumably $\sim 1$ km. This leads to
debris disks that are too bright at early times and that are too dim
at late times, by a couple orders of magnitude.

We do not believe these discrepancies can be resolved by relaxing some
of our model assumptions.  In particular, we argue that our estimate
for $q_3$ is unchanged even taking into account the fact that disk
eccentricity may rise with time. Our results are also insensitive to the width of the debris
ring, to the strength of bodies, and to the assumed upper and lower
sizes.

Because we restrict our attention to the upper envelope of inferred
masses, our result is dominated by a handful of systems. Our analysis
may be vulnerable to errors. However, the evidence is solid that
debris disks remain fairly bright even at a few billion years. This
alone dictates that there ought to be lots of mass stored in large
(10-100 kms) planetesimals. We address how this is accomplished by
revisiting coagulation model in an upcoming publication.

\begin{acknowledgements}
  Y. Wu thanks Y. Lithwick, H. Schlichting and P. Sari for
  discussions. We acknowledge financial support by NSERC.
\end{acknowledgements}

\appendix

\section{ Evolution of Debris Disk Properties: Analytical Model}
\label{sec:simplemodel}

In the following, we present a simple analytical model that describes
the time evolution of dust luminosity and size distribution in debris
disks. This model is very similar to that described in
\citet{Lohne:2008}, except for our choice for the size of the largest
planetesimals. In the following, we present results with arbitrary
strength law and initial size distribution, followed by numerical
evaluations using the hard strength law and for $q_3 = 4$.

We approximate the body strength (eq. \ref{BenzAndAsphaug}) by two
broken power-laws,
\begin{equation}
\bar{Q}_D = 
\begin{cases}
A\left(\frac{s}{1 \rm{cm}}\right)^{\alpha} & \quad s < s_1\\
B\rho \left(\frac{s}{1 \rm{cm}}\right)^{\beta} & \quad s > s_1\\
\end{cases}
\end{equation}
where $s_1$~is the size at which the two expressions meet. The body
strength is dominated by material strength below $s_1$ and by
self-gravity above $s_1$. For the hard strength law that we adopt,
$\alpha = -0.3, \beta=1.5$ and $s_1 =
\left(\frac{B}{A}\rho\right)^{\frac{1}{\alpha-\beta}} \rm{cm} \approx
300$ meters.

Combined with equation \refnew{eq:vimpact}, the minimum size of an
impactor that causes catastrophic disruption is
\begin{equation}
s_{\rm{impactor}} =
\begin{cases}
\left(\frac{2A}{1.75e^2v_{\rm{kep}}^2}\right)^{\frac{1}{3}}\left(\frac{s}{\mbox{1 cm}}\right)^{1+\frac{\alpha}{3}} = k_1s^{\kappa_1} & \quad s < s_1\\
\left(\frac{2B\rho}{1.75e^2v_{\rm{kep}}^2}\right)^{\frac{1}{3}}\left(\frac{s}{\mbox{1 cm}}\right)^{1+\frac{\beta}{3}} = k_2s^{\kappa_2} & \quad s > s_1\\
\end{cases}
\label{eq:simp}
\end{equation}
Here, $\kappa_1 = 0.9$ and $\kappa_2 = 1.5$ for our adopted strength law.

We define a break-size, $s_2 = s_2(t)$, to be the size at which the
time-integrated chance of destruction per body is unity, or the
optical depth for size $s_2$ to be hit is,
\begin{equation}
\tau (s_2) = {{t_{\rm orb}}\over {t}}.
\label{eq:taus2}
\end{equation}  
Bodies larger than $s_2$ have hardly collided and they retain their
primordial size distribution, while bodies smaller than $s_2$ have
collided many times, and they satisfy the size distribution for
collisional equilibrium. If $s_2 > s_1$, we adopt a size distribution
that is piece-wise continuous,
\begin{equation}
\frac{dn}{ds} =
\begin{cases}
n_1s^{-q_1} = n_3s_1^{q_1-q_2}s_2^{q_2-q_3}s^{-q_1}& \quad s < s_1\\
n_2s^{-q_2} = n_3s_2^{q_2-q_3}s^{-q_2} & \quad s_1 < s < s_2\\
n_3s^{-q_3} & \quad s > s_2 \\
\end{cases}
\label{eq:Three}
\end{equation}
where $q_1$ and $q_2$ are the power indexes at collisional
equilibrium.  They are $3.5$ \citep{Dohnanyi:1969} if the size ratio
between the impactor and the target is constant. Given equation
\refnew{eq:whatisq}, we have $q_1 = 3.6$ and $q_2 = 3.0$ for the hard
strength law. This piece-wise size distribution breaks down near the
blow-out size due to an abrupt deficit of small bullets. A more
accurate derivation for the size distribution can be obtained by
assuming that the mass loss rate is constant with size, as is carried
out in \citet{StrubbeChiang}.  The size distribution shows a
  flare-up toward the blow-out size, and the magnitude of the flare-up
  depends on, among other things, the value of eccentricity.  Our analytical results obtained based on equation \refnew{eq:Three}
  should be regarded as illustrative.

We first obtain the evolution of $s_2$ with time.  When $s_2 < s_1$,
i.e., collisions involve only bodies bound by the material strength,
optical depth for destruction at $s_2$ is determined by integrating
over all its possible bullets,
\begin{equation}
\tau(s_2) = \frac{n_3 s_2^{q_1-q_3}}{2\pi a\Delta a}
\int_{k_1s_2^{\kappa_1}}^{s_2}\, \pi \left(s_2 + s\right)^2s^{-q_1}ds =  \frac{n_3}{2a\Delta a}\frac{1}{q_1-1}k_1^{1-q_1} s_2^{q_1-q_3+2+\kappa_1 - \kappa_1 q_1}.
\label{eq:materialgrowth}
\end{equation}
Substituting this into the definition for $s_2$ (eq. \ref{eq:taus2}),
we obtain
\begin{equation}
s_2 \propto t^{1\over{-q_1+q_3-2-\kappa_1+\kappa_1 q_1}}
\label{eq:s2a}
\end{equation}
This yields $s_2 \propto t^{1.4}$ for our parameters.

Once $s_2 > s_1$, we perform the same exercise and obtain,
\begin{equation}
s_2 \propto t^{1\over{-q_2+q_3-2-\kappa_2+\kappa_2 q_2}} 
\label{eq:s2b}
\end{equation}
or $s_2 \propto t^{0.5}$ for our parameters. So at early times, the
break size rises steeply with time, due to an abundance of small
bullets; while at late times, the break size rises with time more
gradually due to the relative paucity of bullets. These two scaling
relations are observed in our numerical results (Figure \ref{fig:s2}).

Now we proceed to derive the scaling of disk luminosity with system
age. We let the infrared luminosity to be that portion of the
starlight that is intercepted by debris particles.  This is directly
related to the total surface area of all particles, which is mostly
contributed by particles around $s_{\rm min}$.\footnote{The upper bound of the integration is chosen to be $s_2$ but it is of no importance.} 
The fractional luminosity is therefore,
\begin{equation}
\frac{L_{\rm{IR}}}{L_*} \approx {{\int_{s_{\rm min}}^{s_2} \pi s^2\, n_1  s^{-q_1} d s }
\over{4 \pi a^2}} \approx 
\begin{cases}
\frac{n_3}{4 a^2 (q_1-3)}s_2^{q_1-q_3}{s_{\rm{min}}^{3-q_1}} 
\quad \quad\quad\quad s_2 < s_1 \\
\frac{n_3}{4 a^2 (q_1-3)}s_1^{q_1-q_2}s_2^{q_2-q_3}
{s_{\rm{min}}^{3-q_1}} \quad \quad s_2 > s_1 \\
\end{cases}
\label{eq:lofs2}
\end{equation}
So the evolution of luminosity is dictated by the evolution of $s_2$
with time. In particular, at late times (when $s_2 > s_1$), the fractional
luminosity decays with time gradually,
\begin{equation}
\frac{L_{\rm{IR}}}{L_*} \propto s_2^{q_2 - q_3} \propto 
 t^{{q_2 - q_3}\over{q_2-q_3+2+\kappa_2-\kappa_2 q_2}}.
\label{eq:ftime}
\end{equation}
Again, for our choice of parameters, $L_{\rm IR}/L_* \propto
t^{-0.5}$.  If $q_3$ alone is varied, $L_{\rm IR}/L_* \propto
t^{(q_3-3)/(2-q_3)}$ and scales as $t^{-1/3}, t^{-1/2}$ and $t^{-2/3}$
for $q_3 = 3.5, 4,$ and $5$ respectively. This forms the basis on
which we decipher the primordial distribution of planetesimals.

To understand the dependence of the fractional luminosity on a range
of parameters, we return to equations \refnew{eq:lofs2},
\refnew{eq:s2b} and \refnew{eq:materialgrowth}, retaining all the
neglected constants and obtaining the following expression,
\begin{equation}
\frac{L_{\rm{IR}}}{L_*} \approx 
\begin{cases}
\frac{n_3}{4 a^2 (q_1-3)}\left[\frac{ 2 a \Delta a}{n_3}\left(q_1-1\right)\frac{t_{\rm{orb}}}{t}\left(\frac{2A}{1.75e^2v_{\rm{kep}}^2}\right)^{\frac{1}{3} \left(q_1-1\right)}\right]^{\frac{q_1-q_3}{2+\kappa_1-\kappa_1 q_1 + q_1 - q_3}}
s_{\rm{min}}^{3-q_1}
\quad\quad\quad\quad\quad\quad s_2 < s_1 \\
\frac{n_3}{4 a^2 (q_1-3)}s_1^{q_1-q_2}
\left[\frac{ 2 a \Delta a}{n_3}\left(q_2-1\right)\frac{t_{\rm{orb}}}{t}\left(
\frac{2B\rho}{1.75e^2v_{\rm{kep}}^2}
\right)^{\frac{1}{3}\left(q_2-1\right)}
\right]^{\frac{q_2-q_3}{2+q_2-q_3+\kappa_2-\kappa_2 q_2}}
s_{\rm{min}}^{3-q_1} \quad\quad\quad s_2 > s_1 \\
\end{cases}
\label{eq:loff}
\end{equation}
Substituting our nominal values for the indexes ($\kappa_1 = 0.9$,
$\kappa_2 = 1.5$, $q_1 = 3.6$, $q_2 = 3.0$, $q_3 = 4.0$), we simplify
the dependency for luminosity into (for at late times when $s_2 > s_1$),
\begin{equation}
{{L_{\rm IR}}\over{L_*}} \propto 
t^{-0.5} M_0^{0.5} a^{-3.6} \left(\frac{\Delta a}{a}\right)^{0.5}
 e^{-\frac{2}{3}} M_*^{\frac{5}{6}}
B^{\frac{1}{3}}A^{-\frac{5}{6}}  s_{\rm{min}}^{-0.6},
\label{eq:frac2}
\end{equation}
where $M_0$ is the total mass of the disk, $a$ its radius, $\Delta
a/a$ its fractional width, $e$ the eccentricity of particles, $M_*$
the central stellar mass, $s_{\rm min}$ the blow-out size, and $A, B$
the strengths.  This relation illuminates how our procedure, using
luminosity to infer $M_0$, can be affected by various parameters.  For
example, the actual position of the belt is a piece of essential
information, while other values should be known roughly to within a
factor of a few to avoid gross mis-estimate.

Equation \refnew{eq:loff} can also be used to
illustrate the effect of a time-varying eccentricity on our estimate
for $q_3$. At a given dust luminosity, the inferred initial mass
scales with the system age and the eccentricity as

\begin{equation}
M_0 \propto t^{q_3-3}e^{\frac{4}{3}\left(q_3-3\right)}.
\label{eq:M0et}
\end{equation}

Let the plantesimals be stirred with a time-dependence of $e \propto
t^\gamma$.  We define a $\bar{q_3}$ as the value of $q_3$ one obtains
by taking a constant eccentricity (in which case $M_0 \propto t^{{\bar
    q_3}-3}$).  The true $q_3$ is related to it as
\begin{equation}
q_3 = 3 + \left({\bar q_3} - 3\right)\left(1+ \frac{4}{3}\gamma\right)^{-1}.
\label{eq:barq3}
\end{equation}
So for $\gamma = 1/4$, ${\bar q_3} = 4$, we get the true $q_3 = 3.75$.

 Numerically we find a weaker dependency on $\gamma$. This is
  related to the afore-mentioned flare-up near the blow-out size. If
  instead of equation \refnew{eq:Three}, we make the simplifying
  assumption that the mass loss rate is the same at blow-out size as
  at other sizes, but that the micron grains are destroyed by similar
  grains (as opposed to smaller ones), we find that the dust
  luminosity is proportional to the total number of blow-out grains,
  while the mass loss rate is proportional to the square of this
  number. As a result, we write
\begin{equation} {{L_{\rm IR}}\over{L_*}} \propto {\dot M}^{1/2}.
\label{eq:LtodMdt}
\end{equation}
From this, we derive the dependence of dust luminosity on time and on
eccentricity that are slightly different from those presented in
equations \refnew{eq:ftime}, \refnew{eq:frac2}, \refnew{eq:M0et} and
\refnew{eq:barq3}. For instance, in contrast to equation
\refnew{eq:barq3}, the dependence of $q_3$ on $\gamma$ is
logarithmic.

\bibliographystyle{apj}
\bibliography{shan0316}

\end{document}